\documentclass[aps,pra,twocolumn,showpacs,superscriptaddress,amsmath,amssymb]{revtex4-1}
\usepackage{graphicx}
\usepackage{epstopdf}
\usepackage{wasysym}
\usepackage{color}

\newcommand{\norm}[1]{\left|\!\left|{#1}\right|\!\right|}
\newcommand{\ket}[1]{\left|{#1}\right>}
\newcommand{\bra}[1]{\left<{#1}\right|}
\newcommand{\tr}[1]{\textnormal{tr}{\left\{#1\right\}}}
\newcommand{\SP}[1]{\textnormal{Sp}{\left\{#1\right\}}}
\newcommand{\TP}[1]{{#1}^\mathrm{\,\textsc{t}}}

\newcommand{\QTTF}[1]{\mathrm{qTTF}\!\left(#1\right)}
\newcommand{\diag}[1]{\mathrm{diag}\!\left(#1\right)}
\newcommand{\RE}[1]{\mathrm{Re}\left\{#1\right\}}

\newcommand{\FMAT}{\pmb{F}}
\newcommand{\CMAT}{\pmb{\mathcal{C}}}
\newcommand{\PMAT}{\pmb{\mathcal{P}}}
\newcommand{\XMAT}{\pmb{\mathcal{X}}}
\newcommand{\YMAT}{\pmb{\mathcal{Y}}}
\newcommand{\OMAT}{\pmb{\mathcal{O}}}
\newcommand{\SMAT}{\pmb{\mathcal{S}}}
\newcommand{\AMAT}{\pmb{\mathcal{A}}}
\newcommand{\BMAT}{\pmb{\mathcal{B}}}

\begin{document}

\title{On determining which quantum measurement performs better for state estimation}

\author{Jaroslav {\v R}eh{\'a}{\v c}ek}
\affiliation{Department of Optics, Palack{\'y} University, 17. listopadu 12, 77146 Olomouc, Czech Republic}
\author{Yong Siah Teo}
\affiliation{Department of Optics, Palack{\'y} University, 17. listopadu 12, 77146 Olomouc, Czech Republic}
\author{Zden{\v e}k Hradil}
\affiliation{Department of Optics, Palack{\'y} University, 17. listopadu 12, 77146 Olomouc, Czech Republic}

\begin{abstract}
We introduce an operational and statistically meaningful measure, the quantum tomographic transfer function, that possesses important physical invariance properties for judging whether a given informationally complete quantum measurement performs better tomographically in quantum-state estimation relative to other informationally complete measurements. This function is independent of the unknown true state of the quantum source, and is directly related to the average optimal tomographic accuracy of an unbiased state estimator for the measurement in the limit of many sampling events. For the experimentally-appealing minimally complete measurements, the transfer function is an extremely simple formula. We also give an explicit expression for this transfer function in terms of an ordered expansion that is readily computable and illustrate its usage with numerical simulations, and its consistency with some known results.
\end{abstract}

\date{\today}

\begin{widetext}
\maketitle
\end{widetext}

Quantum-state estimation constitutes a broad class of tools catered to the verification and diagnostics of states for quantum systems, a necessary step in accessing the validity of quantum protocols. Thus far, much work has been devoted to the development of statistical methods for quantum-state reconstruction \cite{ml,mlme,bayes}, to the investigation on the tomographic accuracies of quantum-state estimators \cite{aj-scott,HZhu1,HZhu2}, to improving the performance of state reconstruction under difficult circumstances (such as large Hilbert-space dimensions, noisy and ill-calibrated detections, \emph{etc.}) with various approaches \cite{scale,data-patt,adaptive,ill-calib}, and to the assignment of statistical measures that quantifies the error and reliability of state estimators \cite{conf-region,ml-region}, and so forth.

This Letter focuses primarily on the comparison of different quantum measurements available for state estimation, which is yet another important aspect in this field. In particular, a value of a certain statistically meaningful quantity is assigned to every quantum measurement that serves as a gauge on their tomographic performance. The ascription of a measurement with a meaningful number that quantifies its performance is not a recent proposal. In optics, one can describe the performance of an optical system with the optical transfer function \cite{OTF}, which incorporates the details of signal propagation from every single component in the optical system to the image plane. In quantum-state estimation, it is also useful to introduce such a function to every quantum measurement.

In quantum-state estimation, the unknown quantum state $\rho$ of a source is reconstructed from a set of measurement data supplied by the quantum measurement apparatus, which is mathematically represented by a probability-operator measure (POM) consisting of $M$ positive operators $\Pi_j$ --- $\sum_j\Pi_j=1$ . For a $D$-dimensional Hilbert space, $\rho$ can always be written as a linear combination of $D^2$ Hermitian, trace-orthonormal basis operators, where one operator is a multiple of the identity $1/\sqrt{D}$, and the rest of the $D^2-1$ operators $\Omega_j$ are traceless. This operator basis conveniently incorporates the unit-trace constraint of the state --- $\tr{\rho}=1$ --- and organizes all the relevant $D^2-1$ independent state parameters to be estimated. The correct probabilities $p_j=\tr{\rho\,\Pi_j}$ for the source and apparatus are related to the state $\rho$ according to Born's rule, which translates to a linear system of equation $\mathbf{p}'=\CMAT\mathbf{t}$, where $\mathbf{t}$ is a column of $D^2-1$ coefficients $\tr{\rho\,\Omega_j}$, $\mathbf{p}'$ has components $p_j-\tr{\Pi_j}/D$, and the central object of our discussion, the $M\times (D^2-1)$ matrix $\CMAT$ with entries $\CMAT_{jk}=\tr{\Pi_j\Omega_k}$, provides complete information about the apparatus. Throughout this discussion, we shall consider only informationally complete POMs ($M\geq D^2$), that is POMs that uniquely characterize $\rho$. A POM is minimally complete if it contains $M=D^2$ outcomes that are all linearly independent.

There exist proposals \cite{kulik-PRL,robust} attempting to judge the efficiency of a quantum tomography protocol by investigating solely the singular values of $\CMAT$. The preferred choice is a measurement corresponding to a $\CMAT$ with the smallest condition number $\kappa=\{\text{ratio of largest to smallest singular values of }\CMAT\}$ out of the conceivable choice. This measurement, according to the proposals, is thus one that maximizes the error stability for the experiment. It can be easily shown that this number is simply inadequate to assess the tomographic accuracies of state estimators. \emph{A word of caution}---The articles \cite{kulik-PRL,robust} compare $M\times D^2$ measurement matrices, which we denote by $\widetilde{\CMAT}$, for the usual linear-inversion (LIN) tomography schemes, rather than comparing the $M\times (D^2-1$) $\CMAT$ matrices discussed here. We point out that the former approach of comparison is incompatible with unit-trace LIN estimators \cite{suppI}. Nevertheless we shall briefly demonstrate the shortcomings arising from taking the condition number $\kappa$ serious using this approach.

Apart from the inability to distinguish the performance between symmetric informationally complete (SIC) POMs ($M=2^2=4$) and mutually unbiased bases (MUB) for $D=2$, for instance, as both give $\kappa=1.7321$, with the latter proven to outperform the former \cite{HZhu2} tomographically, the condition number of $\widetilde{\CMAT}$ has a serious flaw as a tomographic measure --- it is not invariant under channel duplication. To illustrate this, we first write down
\begin{equation}
\widetilde{\CMAT}\,\widehat{=}\begin{pmatrix}
0.3536 &  -0.2041  &  0.2041  &  0.2041\\
0.3536 &   0.2041  &  0.2041  & -0.2041\\
0.3536 &   0.2041  & -0.2041  &  0.2041\\
0.3536 &  -0.2041  & -0.2041  & -0.2041
\end{pmatrix}
\end{equation}
for the qubit SIC~POM with a set of basis operators, whose singular values are $\{0.7071,0.4082,0.4082,0.4082\}$, such that $\kappa=1.7321$. Next, the fourth outcome of this POM (last row of $\widetilde{\CMAT}$), say, is duplicated with equal proportions, so that the resulting $5\times 4$ matrix
\begin{equation}
\widetilde{\CMAT}'\,\widehat{=}\begin{pmatrix}
\cdots & \cdots & \cdots & \cdots\\
0.1768 &  -0.1021  & -0.1021  & -0.1021\\
0.1768 &  -0.1021  & -0.1021  & -0.1021
\end{pmatrix}
\end{equation}
now has singular values  $\{0.6700,0.4082,0.4082,0.3047\}$, with $\kappa=2.1988$. Physically, duplicating any channel(s) cannot affect the tomographic information gain. The condition number $\kappa$, and other functions of the singular values of $\widetilde{\CMAT}$ alone for that matter would, however, report completely nonsensical results.

The appropriate quantity to consider for tomographic-accuracy estimation is the trace of the inverse of the scaled Fisher information matrix $\FMAT(\rho)=\TP{\CMAT}\PMAT^{-1}\CMAT$ for the usual multinomial detection statistics as the meaningful quantity in determining the performance of a POM, with $\PMAT=\diag{p_1,p_2,\ldots,p_M}$. This quantity is well-motivated. Firstly, if the unknown state $\rho$ is full-rank, which is always the case in realistic experiments, the matrix trace $\SP{\FMAT(\rho)^{-1}}$ gives the scaled (over sampling events) optimal tomographic accuracy for all unbiased state estimators $\widehat{\rho}$ in the limit of large sampling events. This accuracy is given as the Hilbert-Schmidt distance between $\rho$ and $\widehat{\rho}$, which is a natural distance for quantum states and has been investigated in, for instance, Ref.~\cite{aj-scott,HZhu1,HZhu2}. Secondly, the scaled Fisher matrix $\FMAT(\rho)$ possesses an important physical invariance property --- it is invariant under channel duplication. Given a POM with $M$ outcomes, no additional information is obtained if any part of this POM is precisely duplicated.

To define a measure that is independent of the unknown state, one can perform an average over the state space to obtain an averaged performance for a given POM. Since the state of the source is unknown, and nearly pure states are typically the quantum states of interest in many quantum protocols, this average may be taken in an uninformative way \cite{krw-jones}, in the sense that the Haar prior for pure states is used for the average. Unfortunately, state averages on $\FMAT(\rho)^{-1}$ are generally difficult to compute. Instead, the average is carried out term-wise on an expansion of the right-hand side of the identity
\begin{equation}
\SP{\FMAT(\rho)^{-1}}=\SP{\overline{\FMAT}^{-1}}+\SP{\dfrac{\XMAT\Delta\PMAT}{\pmb{1}-\YMAT\Delta\PMAT}}
\label{eq:Fseries}
\end{equation}
in powers of $\Delta\PMAT=\PMAT-\overline{\PMAT}$. Here, $\overline{\PMAT}$ is the diagonal matrix of probabilities $\overline{p}_j=\tr{\Pi_j}/D$ for the maximally-mixed state, $\overline{\FMAT}=\FMAT(1/D)$, $\XMAT=\overline{\PMAT}^{-1}\CMAT\overline{\FMAT}^{-2}\TP{\CMAT}\overline{\PMAT}^{-1}$, and $\YMAT=\overline{\PMAT}^{-1}\CMAT\overline{\FMAT}^{-1}\TP{\CMAT}\overline{\PMAT}^{-1}-\overline{\PMAT}^{-1}$. The positive matrices $\XMAT$ and $-\YMAT$ have several interesting properties. First of all, they are mutually conjugate orthogonal --- $\XMAT\overline{\PMAT}\YMAT=\pmb{0}=\YMAT\overline{\PMAT}\XMAT$. The matrix $-\overline{\PMAT}$ turns out to be the generalized inverse of $\YMAT$ --- $\YMAT\overline{\PMAT}\YMAT=-\YMAT$. The matrix $\YMAT$ also belongs to the kernel of $\CMAT$ --- $\TP{\CMAT}\YMAT=\pmb{0}=\YMAT\CMAT$.

A power series for $\SP{\FMAT(\rho)^{-1}}$ exists only when the eigenvalues of $\norm{\YMAT\Delta\PMAT}$ are sufficiently small. It is possible to take advantage of the scaling transformation $\FMAT(\alpha\rho)=\FMAT(\rho)/\alpha$ to obtain convergence series for any $\rho$ by setting $\alpha<\alpha_0=1/\left(\norm{Y}_2\max_j\{{\tr{\Pi_j}}\}\right)$ \cite{suppII}. After taking the Haar average of this series, we define the tomographic measure for any POM as
\begin{align}
&\,\QTTF{\{\Pi_j\}}\equiv\overline{\SP{\FMAT(\rho)^{-1}}}\nonumber\\
=&\,\underbrace{\SP{\overline{\FMAT}^{-1}}}_{\text{zeroth order}}+\underbrace{\dfrac{\alpha}{D(D+1)}\sum^M_{j_1,j_2=1}\XMAT_{j_2j_1}\YMAT_{j_1j_2}\pmb{\mathcal{G}}^{(2)}_{j_1j_2}}_{\text{second order}}+\ldots\,,
\label{eq:Fseries_avg}
\end{align}
where $\pmb{\mathcal{G}}^{(n)}_{j_1j_2\ldots j_n}=\tr{\Pi_{j_1}\Pi_{j_2}\ldots\Pi_{j_n}}$ are elements of high-order Gram matrices, which is the main mathematical result of this Letter \cite{suppIII}. As this function keeps track of all connections related to the quantum measurement with the average optimal tomographic accuracy of an unbiased $\widehat{\rho}$, we shall name it the \emph{quantum tomographic transfer function} (qTTF) for convenience, in a similar spirit as the optical transfer function in classical optics. Terminologies and acronyms aside, one should remember that this function was defined on a solid conceptual framework in statistics.

One can now understand the fallacies arising from solely studying the condition number of $\CMAT$. In its real singular-value decomposition, $\CMAT=\OMAT \SMAT\TP{\OMAT'}$, with $\TP{\OMAT}\OMAT=1=\OMAT'\TP{\OMAT'}=\TP{\OMAT'}\OMAT'$, $\SMAT$ and $\OMAT'$ of dimensions $(D^2-1)\times(D^2-1)$ and $\OMAT$ of dimensions $M\times(D^2-1)$, the quantity $\SP{\FMAT(\rho)^{-1}}=\SP{(\SMAT\TP{\OMAT}\PMAT^{-1}\OMAT \SMAT)^{-1}}$ depends on \emph{both} the diagonal matrix $\SMAT$ of singular values, as well as the orthogonal matrix $\OMAT$. While $\SMAT$ loosely describes the distribution of tomographic weights given to the measurement outcomes, the matrix $\OMAT$ contains information about the mutual relationships among these outcomes. The complete information about the optimal tomographic accuracy of $\widehat{\rho}$ that is carried by these two mathematical objects is distributed throughout the series in Eq.~\eqref{eq:Fseries_avg}, with part of this information manifested in the higher order Gram matrices $\pmb{\mathcal{G}}^{(n)}$. Leaving any of these objects out of the description renders any POM comparison potentially illegitimate. For instance, it is a simple matter to find two POMs such that the one with the bigger $\kappa$ value (weakly conditioned) gives the lower qTTF value (greater optimal tomographic accuracy).

For higher-order terms in Eq.~\eqref{eq:Fseries_avg} determine the difference between the averaged quantity and the quantity evaluated with the averaged state. The formula shows that this difference can only be accounted for by Gram matrices of \emph{all} orders. The comprehensive role of these Gram matrices is now clear --- they contain all crucial mutual relationships among the measurement outcomes that enter the tomographic error propagation to $\widehat{\rho}$, a characteristic that is analogous to the mechanism behind the optical transfer function, which encompasses all relative phase information of the signal through the individual components of an optical system.

There exist an extremely simple formula for the qTTF when evaluating minimally complete POMs. For POMs of this type, since $\YMAT$ resides in the kernel of $\CMAT$, the rank of $\YMAT$ must therefore be one, for $\CMAT$ now has a $(D^2-1)$-dimensional row space. Since $\sum_j\CMAT_{jk}=0$, $\YMAT$ must therefore be a negative matrix with entries all equal to minus one in the computational basis --- $\YMAT_{j,k}=-1$. Then, we have $\YMAT\PMAT\YMAT=-\YMAT$ for any probability matrix $\PMAT$ so that the second term in the right-hand side of Eq.~\eqref{eq:Fseries} has a finite series expansion consisting of only the two lowest-order terms in $\Delta\PMAT$. Thus, Eq.~\eqref{eq:Fseries_avg} with $\alpha=1$ greatly simplifies to
\begin{equation}
\QTTF{\{\Pi_j\}_\textsc{min}}=\SP{\overline{\FMAT}^{-1}_\textsc{min}}-1+\dfrac{1}{D}
\label{eq:Fseries_avg_MIN}
\end{equation}
for any minimally complete POM $\{\Pi_j\}_\textsc{min}$. For overcomplete POMs made up of $D+1$ bases (minimally complete bases) of $D$ rank-one outcomes [$M=D(D+1)$], the qTTF is also completely described by only the first two terms of Eq.~\eqref{eq:Fseries_avg} with $\alpha=1$, as it can be shown that $\YMAT$ is a rank-($D+1$) projector that takes a block-diagonal form, with $D+1$ blocks each of dimensions $D\times D$ and all matrix elements equal to $-(D+1)$ for every block, so that again $\YMAT\PMAT\YMAT=-\YMAT$ and
\begin{equation}
\QTTF{\{\Pi_j\}_\textsc{min bases}}=\dfrac{\SP{(\TP{\CMAT}\CMAT)^{-1}}}{(D+1)^2}\,.
\label{eq:Fseries_avg_MIN_BAS}
\end{equation}

The formulas presented in this Letter can be verified to be consistent with known results in quantum tomography. To this end, we note that the trace of the scaled Fisher matrix $\SP{\overline{\FMAT}}\leq D(D-1)$ evaluated at $\PMAT=\overline{\PMAT}$ is bounded from above by $D(D-1)$ for \emph{any} informationally complete POM. With the help of the inequality $\SP{\pmb{\mathcal{A}}}\SP{\pmb{\mathcal{A}}^{-1}}\geq\left(\text{dim}\!\left\{\pmb{\mathcal{A}}\right\}\right)^2$, which is saturated when $\pmb{\mathcal{A}}$ is a multiple of the identity, we find that the zeroth term in Eq.~\eqref{eq:Fseries_avg} is bounded from below inasmuch as
\begin{equation}
\SP{\overline{\FMAT}^{-1}}\geq\dfrac{(D+1)(D^2-1)}{D}\,,
\label{eq:zeroth_bound}
\end{equation}
which again holds for any informationally complete POM. For minimally complete POMs, the equality in \eqref{eq:zeroth_bound} is attained for SIC~POMs, where $\overline{\FMAT}=\overline{\FMAT}_\textsc{min}=\overline{\FMAT}_\textsc{sic}=D/(D+1)$ is a multiple of the $(D^2-1)$-dimensional identity. We thus obtain the well-known lower bound for the qTTF evaluated with SIC~POMs of dimension $D$ to be
\begin{equation}
\QTTF{\{\Pi_j\}_\textsc{sic}}=D^2+D-2
\label{eq:Fseries_avg_SIC}
\end{equation}
for minimally complete POMs, as reported in Refs.~\cite{aj-scott,HZhu1,HZhu2}. When $D$ is a prime power, it can be shown that any complete set of MUB will also saturate \eqref{eq:zeroth_bound}, and since the inequality $\SP{\pmb{\mathcal{A}}}\leq a_\text{max}\,\text{dim}\!\left\{\pmb{\mathcal{A}}\right\}$ with respect to the largest eigenvalue $a_\text{max}$ of $\pmb{\mathcal{A}}$ is saturated when $\pmb{\mathcal{A}}$ is a multiple of the identity, the right-hand side of Eq.~\eqref{eq:Fseries_avg_MIN_BAS} is correspondingly minimized to
\begin{equation}
\QTTF{\{\Pi_j\}_\textsc{mub}}=D^2-1<\QTTF{\{\Pi_j\}_\textsc{sic}}
\label{eq:Fseries_avg_MUB}
\end{equation}
over all minimally complete bases \cite{HZhu2}.

For any other kinds of overcomplete POMs, there is no simple closed-form expression for the series of qTTF in Eq.~\eqref{eq:Fseries_avg}. Although it is straightforward to obtain higher order terms as desired, the computation becomes exponentially exhaustive as either $M$ or $D$ increases. We shall now highlight the viable procedures for numerically computing the first few computable terms of the qTTF for different regimes of $M$ and $D$. If both $M>D^2$ and $D$ are not too large, then the qTTF of any POM can be found by computing higher-order corrections in Eq.~\eqref{eq:Fseries_avg} for $\alpha\approx\alpha_0$. If $D$ is large, so that the computation of high-order terms starts to become expensive, and yet $M$ is not too far from $D^2$, it turns out that taking the sum of the zeroth- and second-order terms for $\alpha=1$ gives good approximation to the qTTF. The largest relative error --- the ratio of the difference between this approximation and the actual qTTF to the actual qTTF --- as $M$ approaches infinity is given by $D/(D+2)$ \cite{suppIV}. One may also take $\alpha\approx\alpha_0$ and attempt to approximate the series with various models, but numerical experience indicates that terminating the series up to the second-order correction for $\alpha=1$ yields more accurate approximations in this regime of $M$ and $D$. If both $M$ and $D$ are large, then performing a Monte Carlo calculation by averaging $\SP{\FMAT(\rho)^{-1}}$ over a Haar ensemble of pure states is the most economical way of computing the qTTF.

\begin{figure}[t]
\centering
\includegraphics[width=1.02\columnwidth]{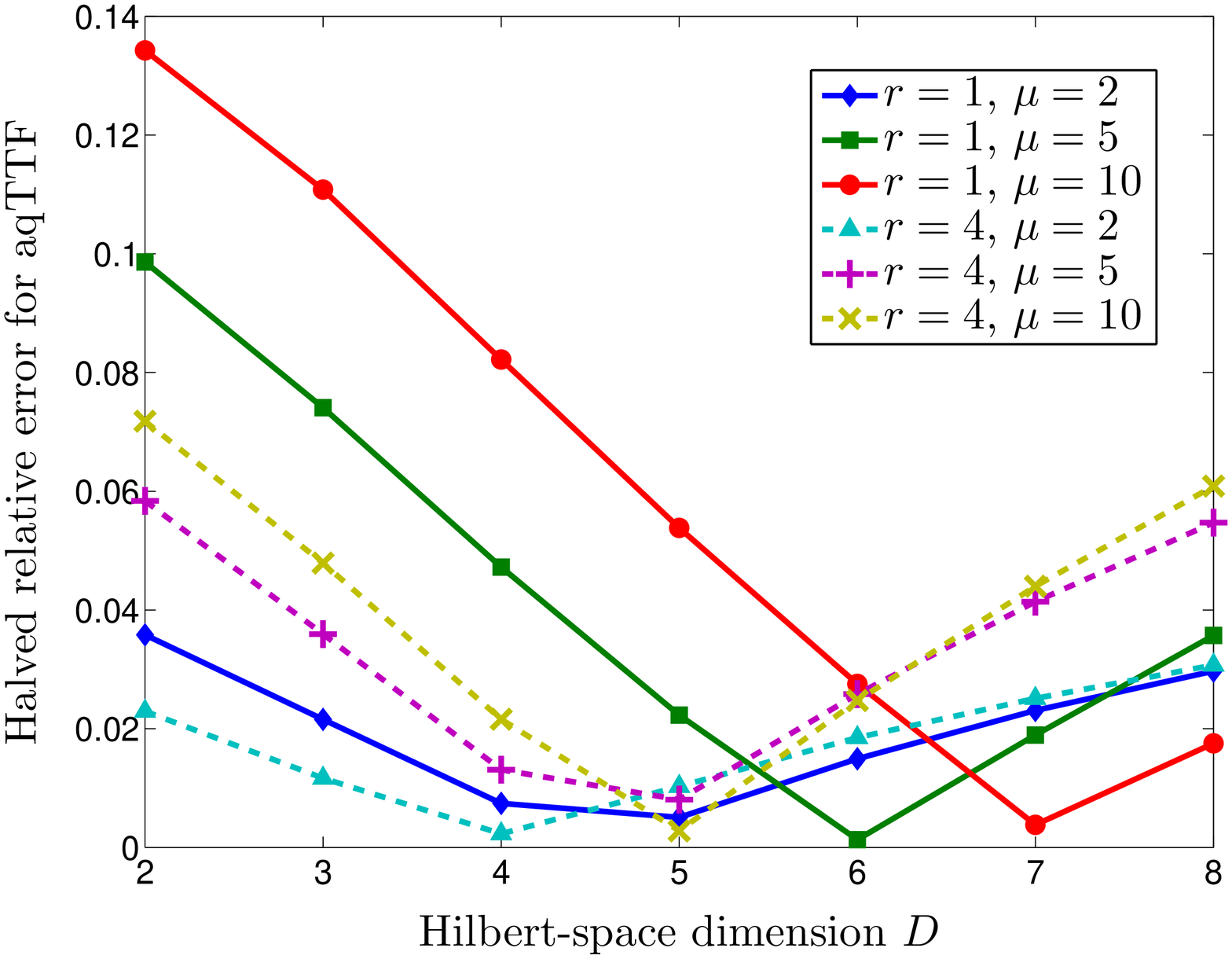}
\caption{(Color online) A plot showing the halved relative error of the aqTTF compared with the correct qTTF value. An average over 500 random rank-one ($r=1$) and full-rank ($r=4$) POMs is carried out separately to compute the data points for different $\mu$ and $D$. The full-rank POMs are generated by taking a convex sum of rank-one POM outcomes and the maximally-mixed state with small admixtures ($\Pi_j=\mathcal{N}_j\{\ket{\psi_j}\bra{\psi_j}+0.05/D\}$) with proper POM normalization $\mathcal{N}_j$. To perform Monte Carlo computations of the qTTF, a Haar set of 500 random pure states is used. For values of $\mu$ for which the number of POM outcomes $M=\mu D^2$ is reasonable, such as the ones shown in the plot, the halved relative error decreases with increasing dimension $D$. The plots indicate that the performance of the second-order correction improves significantly for the slightly mixed POM.}
\label{fig:fig1}
\end{figure}

\begin{figure}[t]
\centering
\includegraphics[width=1.0\columnwidth]{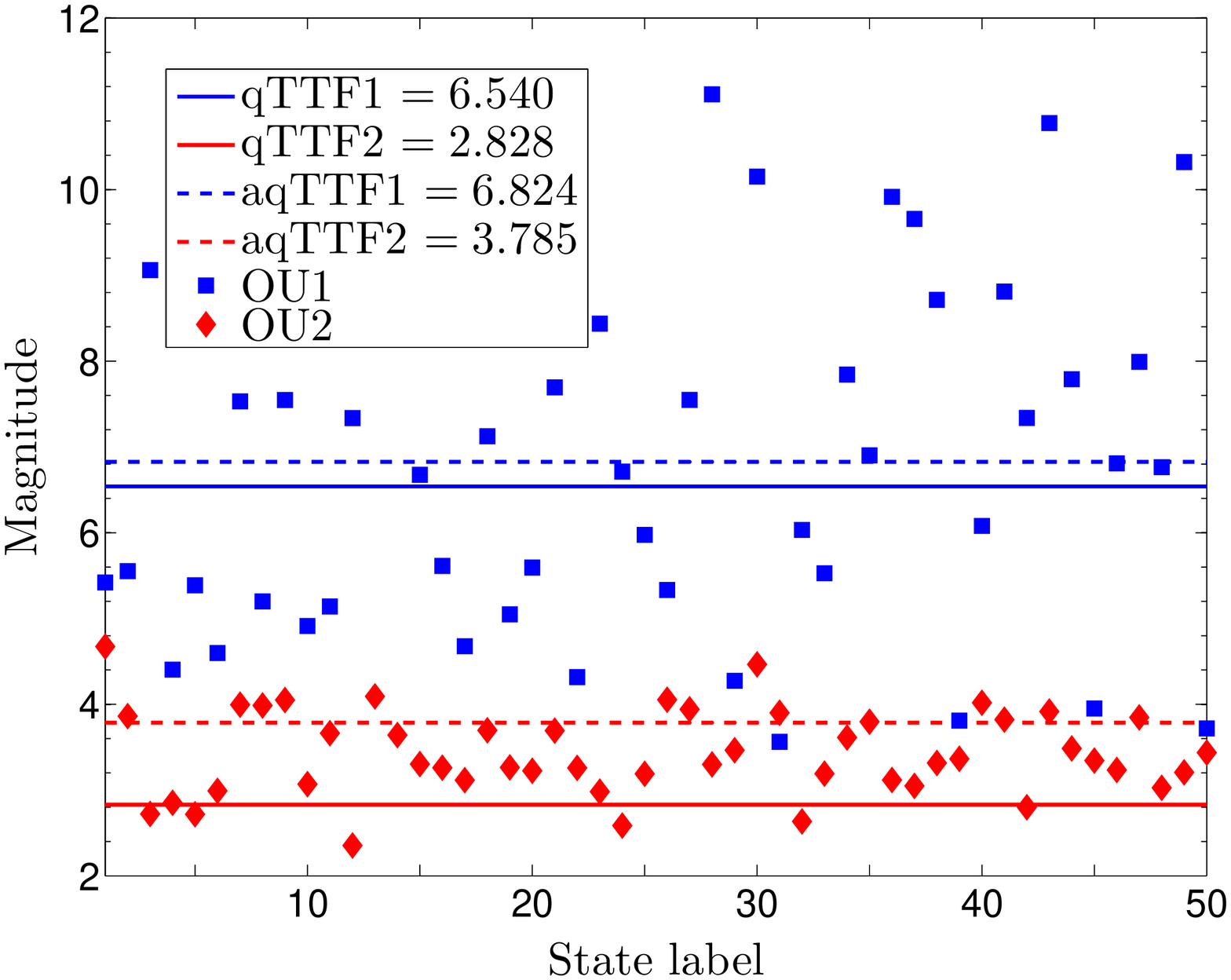}
\caption{(Color online) Comparison of different tomographic quantifiers for two chosen qubit ($D=2$) overcomplete POMs of rank-one, with POM~1 associated with $\kappa_1=3.437$ and POM~2 associated with $\kappa_2=8.119$. A Haar set of 500 random pure states are used for computing the respective qTTFs for both POMs. For both POMs, the averaged Hilbert-Schmidt distance between $\widehat{\rho}$ and $\rho$ is evaluated with $50$ random states, each of which is a pure state that is slightly mixed with the maximally-mixed state, so that the purity is fixed at $\approx0.990$. It is clear that the average tomographic performance of the OU estimator over all the random states lies near the qTTF and aqTTF for each POM, of course, which tells us that POM~2 performs much better than POM~1 for nearly pure states, whereas a contradictorily wrong conclusion would otherwise have been drawn by naively comparing $\kappa_1$ and $\kappa_2>\kappa_1$. ML estimators (not shown) were also computed, and their average performance also lie closely to the respective qTTFs and aqTTFs, as they should.}
\label{fig:fig2}
\end{figure}

We supply two figures to illustrate the validity of qTTF and its calculation procedures. Figure~\ref{fig:fig1} shows, for various $D$, the halved relative error of the second-order approximation of qTTF~(aqTTF) compared with the correct qTTF value (averaged over many random POMs) obtained by averaging $\SP{\FMAT(\rho)^{-1}}$ over a set of random pure states distributed according to the Haar measure \cite{Haar}. The one-half factor originates from error propagation to $\widehat{\rho}$. We generate random POMs for averaging that mimic those used in real experimental scenarios. A set of $M$ complex positive operators $B_j=A_j^\dagger A_j/\tr{A_j^\dagger A_j}$ is first generated, with the $\text{rank}\{\Pi_j\}\times D$ matrices $A_j$ having complex entries distributed according to the standard Gaussian distribution. The resulting POM is obtained with its outcomes given by $\Pi_j=(\sum_jB_j)^{-1/2}B_j(\sum_jB_j)^{-1/2}$ that sum to the identity. The set of random outcomes generated this way tend to have similar traces and this models the typical measurement outcomes employed in an experiment, where the slight variation in the traces originate from systematic instrumental errors and losses that result in non-unit detection efficiencies. Figure~\ref{fig:fig1} tells us that for moderate $M$, the second-order approximation for Eq.~\eqref{eq:Fseries_avg} indeed works rather well even for large $D$. For extremely large values of $M$, the halved relative errors approach the limiting value $D/[2(D+2)]$ for rank-one POMs. Witnessing this limit is, however, a rather impractical feat in any experiment. Moreover, in a realistic situation, the POM outcomes designed may have small amounts of white noise, and the aqTTF is significantly more accurate. As the amount of white noise increases, the zeroth-order term gets increasingly more accurate and all other correction terms vanish since $\Delta\PMAT\approx \mathbf{0}$.

Figure~\ref{fig:fig2} compares the values of four quantities: the condition number $\kappa$ for $\widetilde{\CMAT}'$, the qTTF, the aqTTF, and the scaled mean squared-error (MSE) of the optimal unbiased (OU) reconstruction scheme \cite{remark_lin}. Comparisons are made between a fixed pair of POMs for each of the different $D$ values. This figure gives counterexamples that confirm, once and for all, that the qTTF is the appropriate quantity for estimating the tomographic accuracy for $\widehat{\rho}$, not $\kappa$.

To conclude, we have emphasized the importance of meaningful statistical quantification of a quantum measurement with the quantum tomographic transfer function, which is based on the Haar average of the trace of the inverse Fisher matrix that is equivalent to the optimal tomographic accuracy for unbiased state estimators, as long as the unknown state of the quantum source is not rank-deficient, which is the case in any experiment. This transfer function can be thought of as a quantum analog of the optical transfer function for optical systems. We gave an explicit expression for this transfer function as a series that is readily computable, and provided numerical evidence for the validity of the transfer function in terms of tomographic accuracy estimation of measurements. This function possesses physical invariance properties that are crucial in properly judging the quality of the measurement. Typically, one can take the second-order approximation of the transfer function as a good approximation as long as the Hilbert-space dimension and the number of outcomes are not too large, and in cases where the known measurement outcomes are full-rank due to slight perturbations by white noise, this approximation improves in accuracy.

The authors would like to thank Huangjun~Zhu for the insightful discussions on the subject of overcomplete tomography. This work is co-financed by the European Social Fund and the state budget of the Czech Republic, project~No.~CZ.1.07/2.3.00/30.0004 (POST-UP), and supported by the Czech Technology Agency, project~No.~TE01020229.

\appendix

\section{Linear-inversion tomography}
\label{secA}
The usual linear-inversion tomography schemes involve optimizing some figures of merit of interest. A very common approach is the minimization of the squared error \mbox{$S=|{\bf f}'-\CMAT{\bf t}|^2$}, where $\bf{f}'$ is a column of components $f_j-\tr{\Pi_j}/D$ and $f_j$ are the frequency data for the POM outcomes obtained from the experiment. It is well-known that $\bf{t}=\CMAT^-\bf{f}$ is the solution to the least-squares problem for this quantity, where $\CMAT^-$ is the Moore-Penrose pseudo-inverse of $\CMAT$ such that $\CMAT^-\CMAT=1$, so that the state estimator $\widehat{\rho}=1/D+\sum_jt_j\Omega_j$. Therefore, it is meaningful to directly compare $\CMAT$ of different POMs.

In the other two proposals, the measurement matrices $\widetilde{\CMAT}$ of dimensions $M\times D^2$ are compared instead. The corresponding least-squares problem involves minimizing the same quantity $S$ introduced in the previous paragraph. It is easy to see that the Moore-Penrose pseudo-inverses of these matrices do not guarantee unit-trace state estimators, and that further enforcing the unit-trace constraint will result in the new estimators being nonlinear in $\widetilde{\CMAT}^-$. The least-squares estimators can in fact be shown to be
\begin{align}
\widehat{\rho}&=\sum^{D^2}_{k=1}\left[\widetilde{\CMAT}^-+\chi\left(\TP{\widetilde{\CMAT}}\widetilde{\CMAT}\right)^{-1}\begin{pmatrix}
{\bf e}\,\,{\bf e}\,\,\cdots\,\,{\bf e}
\end{pmatrix}\right]_{k}\Gamma_k\,,\nonumber\\
\chi&=\dfrac{1-\sqrt{D}\,\TP{{\bf e}}\left(\TP{\widetilde{\CMAT}}\widetilde{\CMAT}\right)^{-1}\TP{\widetilde{\CMAT}}{\bf f}}{\sqrt{D}\,\TP{{\bf e}}\left(\TP{\widetilde{\CMAT}}\widetilde{\CMAT}\right)^{-1}{\bf e}}\,,
\end{align}
where $\Gamma_k$ form a complete set of $D^2$ Hermitian trace-orthonormal basis operators and $\TP{{\bf e}}\widehat{=}(1,0,\ldots,0)$. The linear-inversion estimators are thus complicated functions of these measurement matrices, and direct comparisons of $\widetilde{\CMAT}$ do not compatibly correspond to the assessment of the quality of these estimators.

The straightforward reason is that the unit-trace constraint is never taken into account automatically in minimizing $S$. Rather, this constraint is to be additionally fulfilled, for instance, by optimizing only the relevant $D^2-1$ independent state parameters that are associated to traceless basis operators, as shown in the main article.

\section{Convergence of power series}
\label{secB}
There exists a power series for the left-hand side of
\begin{equation}
\SP{\FMAT(\rho)^{-1}}=\dfrac{1}{\alpha}\left[\SP{\overline{\FMAT}^{-1}}+\SP{\dfrac{\XMAT\left(\alpha\PMAT-\overline{\PMAT}\right)}{\pmb{1}-\YMAT\left(\alpha\PMAT-\overline{\PMAT}\right)}}\right]\,,
\end{equation}
which is just a scaling transformation of Eq.~\eqref{eq:Fseries} in the main Letter, as long as the largest eigenvalue of $\YMAT\left(\alpha\PMAT-\overline{\PMAT}\right)$, denoted by $\sigma_\text{max}\!\left\{\YMAT\left(\alpha\PMAT-\overline{\PMAT}\right)\right\}$, is small. Using the inequalities
\begin{align}
\sigma_\text{max}\{\AMAT-\BMAT\}\leq&\,\max\{\sigma_\text{max}\{\AMAT\},\sigma_\text{max}\{\BMAT\}\}\,,\nonumber\\ \sigma_\text{max}\{\AMAT\BMAT\}\leq&\,\sigma_\text{max}\{\AMAT\}\sigma_\text{max}\{\BMAT\}
\end{align}
for any positive matrices $\AMAT$ and $\BMAT$, it can be shown that $\alpha<1/\left(\norm{Y}_2\max_j\{{\tr{\Pi_j}}\}\right)$ is a sufficient condition for this existence.

\section{Power series}
\label{secC}
By noting that the Haar average
\begin{equation}
\overline{\left(\ket{\,\,\,}\bra{\,\,\,}\right)^{\otimes n}}=\dfrac{S_n}{\tr{S_n}}
\end{equation}
over all pure statistical operators of a given Hilbert-space dimension $D$ is related to the permutation projector $S_n$ on the $n$-fold $D$-dimensional symmetric subspace, it is possible to derive the series
\begin{widetext}
{\allowdisplaybreaks
\begin{align}
&\,\QTTF{\{\Pi_j\}}=\underbrace{\SP{\overline{\FMAT}^{-1}}}_{\text{zeroth order}}+\underbrace{\alpha F_2}_{\text{second order}}+\underbrace{\alpha^2(F_3-F_2)+\alpha F_2}_{\text{third
order}}+\underbrace{\alpha^3(F_4-2F_3+F_2)+2\alpha^2(F_3-F_2)+\alpha F_2}_{\text{fourth order}}+\ldots\,,
\label{eq:Fseries_avg}
\end{align}
where
\begin{align}
F_2&=\dfrac{1}{D(D+1)}\sum^M_{j_1,j_2=1}\XMAT_{j_2j_1}\YMAT_{j_1j_2}\pmb{\mathcal{G}}^{(2)}_{j_1j_2}\,,\nonumber\\
F_3&=\dfrac{2}{D(D+1)(D+2)}\left[\sum^M_{j_1,j_2=1}\XMAT_{j_2j_1}\YMAT_{j_1j_2}\pmb{\mathcal{G}}^{(2)}_{j_1j_2}+\!\!\!\sum^M_{j_1,j_2,j_3=1}\!\!\!\XMAT_{j_3j_1}\YMAT_{j_1j_2}\YMAT_{j_2j_3}\RE{\pmb{\mathcal{G}}^{(3)}_{j_1j_2j_3}}\right]\,,\nonumber\\
F_4&=\dfrac{1}{D(D+1)(D+2)(D+3)}\left[\vphantom{\sum^M_{j_1,j_2=1}}\right.(6-D)\sum^M_{j_1,j_2=1}\XMAT_{j_2j_1}\YMAT_{j_1j_2}\pmb{\mathcal{G}}^{(2)}_{j_1j_2}+12\!\!\!\sum^M_{j_1,j_2,j_3=1}\!\!\!\XMAT_{j_3j_1}\YMAT_{j_1j_2}\YMAT_{j_2j_3}\RE{\pmb{\mathcal{G}}^{(3)}_{j_1j_2j_3}}\nonumber\\
&\,+\left.\!\!\!\!\!\!\sum^M_{j_1,j_2,j_3,j_4=1}\!\!\!\!\!\XMAT_{j_4j_1}\YMAT_{j_1j_2}\YMAT_{j_2j_3}\YMAT_{j_3j_4}\left(2\,\RE{\pmb{\mathcal{G}}^{(4)}_{j_1j_2j_3j_4}+\pmb{\mathcal{G}}^{(4)}_{j_1j_2j_4j_3}+\pmb{\mathcal{G}}^{(4)}_{j_1j_3j_2j_4}}+\pmb{\mathcal{G}}^{(2)}_{j_1j_2}\pmb{\mathcal{G}}^{(2)}_{j_3j_4}+\pmb{\mathcal{G}}^{(2)}_{j_1j_3}\pmb{\mathcal{G}}^{(2)}_{j_2j_4}+\pmb{\mathcal{G}}^{(2)}_{j_1j_4}\pmb{\mathcal{G}}^{(2)}_{j_2j_3}\right)\right]\,.
\label{eq:Fseries_avg}
\end{align}
}
\end{widetext}

\section{Second-order correction term}
\label{secD}
For any nonadaptive quantum-state tomography scheme using a fixed POM of $M$ outcomes, the limiting tomographic performance as $M$ approaches infinity is defined by that of the covariant measurement for a given Hilbert-space dimension $D$~[6], whose outcomes themselves form a Haar ensemble of pure states. Therefore, the relative error between the approximate qTTF, obtained \emph{via} the second-order correction, and the actual qTTF approaches the upper bound defined by this measurement.

To get this upper bound, it is enough to argue that the zeroth-order term gives precisely the minimum value $(D+1)(D^2-1)/D$ and show that the difference between the approximate qTTF and the actual value is $4(D^2-1)/(D+2)$. Since the qTTF for the covariant measurement is $2(D-1)$, the relative error
is given by $D/(D+2)$.


\begin{thebibliography}{99}
\bibitem{ml}
M.~Paris and J.~{\v R}eh{\'a}{\v c}ek, \textit{Lecture Notes in Physics --- Quantum State Estimation\/} (Springer, Berlin Heidelberg 2004); J.~{\v R}eh{\'a}{\v c}ek, Z.~Hradil, E.~Knill, and A.~I.~Lvovsky, \pra~\textbf{75}, 042108 (2007).
\bibitem{mlme}
A.~R.~Rossi and M.~G.~A.~Paris, Eur.~Phys.~J.~D~\textbf{32}, 223~(2005); Y.~S.~Teo, H.~Zhu, B.-G. Englert, J.~{\v R}eh{\'a}{\v c}ek, and Z.~Hradil, \prl~\textbf{107}, 020404 (2011); Y.~S.~Teo, B.-G.~Englert, J.~{\v R}eh{\'a}{\v c}ek, and Z.~Hradil, \pra~\textbf{84}, 062125 (2011).
\bibitem{bayes}
R.~Schack, T.~A.~Brun, and C.~M.~Caves, \pra~\textbf{64}, 014305 (2001); R.~Blume-Kohout and P.~Hayden, \texttt{eprint} arXiv:0603116 [quant-ph] (2006); R.~ Blume-Kohout, New~J.~Phys.~\textbf{12}, 043034 (2010); V.~S.~Shchesnovich, \texttt{eprint} arXiv:1403.5158 (2014).
\bibitem{aj-scott}
A.~J.~Scott, J.~Phys.~A~\textbf{39}, 13507 (2006).
\bibitem{HZhu1}
H.~Zhu and B.-G.~Englert, \pra~\textbf{84}, 022327 (2011).
\bibitem{HZhu2}
H.~Zhu, \pra~\textbf{90}, 012115 (2014).
\bibitem{scale}
S.~T.~Flammia, D.~Gross, Y.-K.~Liu, and J.~Eisert, New~J.~Phys.~\textbf{14}, 095022 (2012); T.~Baumgratz, D.~Gross, M.~Cramer, and M.~B.~Plenio, \prl~\textbf{111}, 020401 (2013); C.~Schwemmer, G.~Toth, A.~Niggebaum, T.~Moroder, D.~Gross, O.~G{\"u}hne, and H.~Weinfurter, \prl~\textbf{113}, 040503 (2014);
\bibitem{data-patt}
J.~{\v R}eh{\'a}{\v c}ek, D.~Mogilevtsev, and Z.~Hradil, \prl~\textbf{105}, 010402 (2010); D.~Mogilevtsev, A.~Ignatenko, A.~Maloshtan, B.~Stoklasa, J.~{\v R}eh{\'a}{\v c}ek, and Z.~Hradil, New~J.~Phys.~\textbf{15}, 025038 (2013); L.~Mo{\v t}ka, B.~Stoklasa, J.~{\v R}eh{\'a}{\v c}ek, Z.~Hradil, V.~Kar{\'a}sek, D.~Mogilevtsev, G.~Harder, C.~Silberhorn, and L.~L.~S{\'a}nchez-Soto, \pra~\textbf{89}, 054102 (2014).
\bibitem{adaptive}
F.~Husz{\'a}r and N.~M.~T.~Houlsby, \pra~\textbf{85}, 052120 (2012); K.~Kravtsov, S.~Straupe, I.~Radchenko, N.~M.~T.~Houlsby, and S.~Kulik, \pra~\textbf{87}, 062122 (2013); D.~H.~Mahler, L.~A.~Rozema, A.~Darabi, C.~Ferrie, R.~Blume-Kohout, and A.~M.~Steinberg, \prl~\textbf{111}, 183601 (2013).
\bibitem{ill-calib}
D.~Mogilevtsev, J.~{\v R}eh{\'a}{\v c}ek, and Z.~Hradil, New~J.~Phys.~\textbf{14}, 095001 (2012); S.~Straupe, D.~Ivanov, A.~Kalinkin, I.~Bobrov, S.~P.~Kulik, and D.~Mogilevtsev, \pra~\textbf{87}, 042109 (2013); Y.~S.~Teo, J.~{\v R}eh{\'a}{\v c}ek, and Z.~Hradil, \pra~\textbf{88}, 022111 (2013).
\bibitem{conf-region}
R.~Blume-Kohout, \texttt{eprint} arXiv:1202.5270 (2006); M.~Christandl and R.~Renner, \prl~\textbf{109}, 120403 (2012), \prl~\textbf{109}, 159903 (2012).
\bibitem{ml-region}
J.~Shang, H.~K.~Ng, A.~Sehrawat, X.~Li, and B.-G.~Englert, New~J.~Phys.~\textbf{15}, 123026 (2013).
\bibitem{OTF}
J.~W.~Goodman, {\it Introduction to Fourier Optics, Third Edition}, Roberts and Company Publishers~(2004).
\bibitem{kulik-PRL}
Y.~I.~Bogdanov, G.~Brida, M.~Genovese, S.~P.~Kulik, E.~V.~Moreva, and A.~P.~Shurupov, \prl~\textbf{105}, 010404 (2010).
\bibitem{robust}
A.~Miranowicz, K.~Bartkiewicz, J.~Pe{\v r}ina~Jr., M.~Koashi, N.~Imoto, and F.~Nori, \texttt{eprint} arXiv:1409.4622 (2014).
\bibitem{krw-jones}
K.~R.~W.~Jones, Ann.~Phys.~\textbf{207}, 140 (1991); K.~R.~W.~Jones, \pra~\textbf{50}, 3682 (1994).
\bibitem{suppI}
See Sec.~\ref{secA} for a brief explanation for this remark.
\bibitem{suppII}
See Sec.~\ref{secB} for a reasoning to arrive at this sufficient criterion for convergence.
\bibitem{suppIII}
See Sec.~\ref{secC} for the series up to the fourth-order correction term.
\bibitem{suppIV}
See Sec.~\ref{secD} for a brief argument for this upper bound.
\bibitem{Haar}
K.~{\.Z}yczkowski and H.-J.~Sommers, J.~Phys.~A: Math.~Gen. \textbf{34}, 7111 (2001).
\bibitem{remark_lin}
Linear-inversion estimators do not minimize the Hilbert-Schmidt MSE measure. The authors would like to refer the reader to Eqs.~(6) and (7) in \cite{HZhu2} for the formulas that lead to OU state estimators with respect to this measure.
%
\end{thebibliography}
\end{document}